\documentstyle[12pt,aps,prd,tighten,preprint,newlfont]{revtex}
\begin{document}

\title{ Early years of neutrino oscillations}
\author{S.M.Bilenky}
\address{Joint Institute for Nuclear Research, Dubna, Russia, and\\
INFN, Sezione di Torino, and Dipartimento di Fisica Teorica, Universit\'a
di Torino,\\via P.Giuria 1,I--10125 Torino, Italy\\
E-mail:bilenky@infn.to.it}
\maketitle

\begin{abstract}
The first papers on neutrino oscillations are shortly reviewed
\end{abstract}
\section{ B. Pontecorvo 1957}

The idea of neutrino oscillations was put forward in Dubna by B.
Pontecorvo. He
came to this idea as early as 1957, soon after parity violation in
$\beta$-decay was discovered by Wu et. al.~\cite{WU:57} and the theory
of two-component neutrino was proposed by Landau~\cite{LA:57}, Lee and
Yang~\cite{Lee:57} and Salam~\cite{SA:57}. Only one type of  neutrinos
was known at that time.

B. Pontecorvo mentioned the possibility of $neutrino \leftrightarrow
antineutrino$ transitions in vacuum for the first time in his paper~
\cite{PO:57} on $muonium \leftrightarrow antimuonium$ transitions (
$\mu^-
e^+ \leftrightarrow \mu^+ e^-$). He looked in the lepton world for
a phenomenon that would be analogous to $K^0 \leftrightarrow
\overline{K}{^0}$ oscillations: 

\begin{quote}
``The possible $K^0$ and $\overline{K}{^0}$ transition, which is due to
the weak interactions, leads to the necessity of considering neutral
$K$-mesons as a superposition of particles $K^0_1$ and $K^0_2$ having a
different combined parity. In the present note the question is treated
whether there exist other {\em mixed} neutral particles (not necessarily
{\em elementary}) besides the $K^0$-mesons, which differ from their
anti-particles and for which the $particle \rightarrow antiparticle$
transitions are not strictly forbidden.''
\end{quote}

At that time (and for many years later) neutrino was generally
believed to be a massless particle. This belief was grounded on the
experimental upper limit of the neutrino mass which was known to be much
smaller  than the mass of the electron
(about 250 eV at that time) and
on the success of the theory of the two-component neutrino that was based 
on the assumption of  a massless neutrino. In the paper~\cite{PO:57} 
B. Pontecorvo wrote:

\begin{quote}
``If the two-component neutrino theory should turn out to be incorrect
(which at present seems to be rather improbable)
and if the conservation law of neutrino charge would not apply,
then in principle $neutrino \leftrightarrow antineutrino$ transitions
could take place in vacuo.''
\end{quote}

B. Pontecorvo wrote his first paper on neutrino oscillations later in
1957~\cite{PO2:57}. At that time F. Reines and C. Cowan~\cite{RECO:57}
were doing the famous experiment which led them to discover the {\em
neutrino} through the observation of the inverse $\beta$-process

\begin{equation}
\overline{\nu}_e +p \rightarrow e^+ + n
\end{equation}

\noindent
induced by reactor antineutrinos.

At the same time R. Davis was also doing an experiment on reactor
antineutrino beam~\cite{DA:57}. R.Davis searched for production of~
$^{37}\!\!Ar$ in
the process

\begin{equation}
\overline{\nu}_e + \, ^{37}\!Cl \rightarrow e^- + \, ^{37}\!\!Ar
\end{equation}

\noindent
that is allowed only if lepton number is not conserved.
A rumour reached B. Pontecorvo that Davis had seen some events (2).
B.Pontecorvo who had earlier been thinking 
about possible $antineutrino
\rightarrow neutrino$ transitions decided to study this possibility
in details in connection with reactor experiments:

\begin{quote}
``Recently the question was discussed~\cite{PO:57}  whether there
exist other
{\em mixed} neutral particles beside the $K^0$ mesons, i.e., particles
that differ from the corresponding antiparticles, with the transitions
between particle and antiparticle states not being strictly forbidden.
It was noted that the neutrino might be such a mixed particle, and
consequently there exists the possibility of real $neutrino
\leftrightarrow antineutrino$ transitions in vacuum, provided that
lepton (neutrino) charge is not conserved. In the present note we make a
more detailed study of this possibility, in which interest has been
renewed owing to recent experiments dealing with inverse beta
processes.''
\end{quote}

B. Pontecorvo assumed that

\begin{quote}
\hspace{1em} (a) the neutrino ($\nu$) and antineutrino
($\overline{\nu}$) that are produced in the processes $p \rightarrow n +
\beta^+ +
\nu$ and $n\rightarrow p + \beta^- + \overline{\nu}$ are not identical
particles;

\hspace{1em} (b) lepton number is not strictly conserved 

\end{quote}

\noindent
and went on:

\begin{quote}
``It follows from the above assumptions that in vacuum a neutrino can be
transformed into an antineutrino and vice versa. This means that the
neutrino and antineutrino are {\em mixed} particles, i.e., a symmetric and
antisymmetric combination of two truly neutral Majorana particles $\nu_1$
and $\nu_2$ of different combined parity\footnote {i.e. CP values}.''
\end{quote}

B. Pontecorvo wrote in the paper~\cite{PO2:57}  that he has
considered this
possibility

\begin{quote}
``\ldots~since it leads to consequences which, in principle, can be
tested experimentally. Thus, for example, a beam of neutral leptons
consisting mainly of antineutrinos when emitted from a nuclear reactor,
will consist at some distance $R$ from the reactor of half neutrinos and
half antineutrinos. Under the condition that $R \leq 1m$~\footnote{
This inequality was based on B. Pontecorvo's first estimate of the
oscillation length of reactor antineutrinos}  neutrino
experiments, reminiscent of the Pais-Piccioni experiments with $K^0$
mesons, become possible. Thus, in the experiment of Cowan and Reines, if
$R \leq 1m$, when the neutral particles from the reactor are
captured by hydrogen, the cross section for formation of neutrons and
positrons should be smaller than the cross section expected on simple
thermodynamic grounds\footnote{B. Pontecorvo had in mind
thermodynamic and dimensional arguments by Bethe and Peierls  
that
allowed to connect the cross section of the inverse $\beta$-decay
processes with the probability of $\beta$-decay.} This is due to the
fact that the beam of neutral leptons, which at creation has a known
probability for initiating the reaction, changes its composition 
on
the way from the reactor to the detector.''
\end{quote}

In this first paper, that was written at the time the two-component
neutrino theory had just appeared and the Davis reactor experiment was not
yet
finished, B. Pontecorvo discussed a possible explanation of the Davis
events:

\begin{quote}
``On the other hand, it is difficult to anticipate the effect of real
$antineutrino \rightarrow neutrino$ transitions in the Davis experiment,
since in this case one deals with non- strictly inverse $\beta$ process
\ldots''
\end{quote}

Later B. Pontecorvo understood that in the framework of the two-component
neutrino theory, fully established after the M. Goldhaber et al.
experiment~\cite{GGS}, right-handed antineutrinos can be transfered in
vacuum
only into sterile right-handed neutrinos that do not take part in
weak interaction processes and cannot induce the process
(2). B. Pontecorvo was in fact the first who introduced in 1967 
the notion of sterile neutrinos so popular nowadays (see later).

\section{ Z. Maki, M. Nakagawa, S. Sakata 1962}

The mixing of two neutrinos was introduced  by
Maki, Nakagawa and Sakata in  1962~\cite{MA}
in the framework of the Nagoya model. In this model nucleons were
considered as bound states of a new sort of matter $B^+$ and leptons.
In addition to usual neutrinos $\nu_e$ and $\nu_\mu$, that they called
weak neutrinos, the authors introduced massive neutrinos $\nu_1$ and
$\nu_2$, that they called true neutrinos (the proton was considered as a
bound state of a $B^+$ and a true neutrino $\nu_1$ etc.). They assumed
that the fields of weak neutrinos and true neutrinos were connected by
an orthogonal transformation:

\begin{quote}
``It should be stressed at this stage that the definition of the
particle state of the neutrino is quite arbitrary; we can speak of {\em
neutrinos} which are different from weak neutrinos but are expressed by
a linear combination of the latter. We assume that there exists a
representation which defines the true neutrinos through some orthogonal
transformation applied to the representation of weak neutrinos:

\begin{equation}
\begin{array}{lll}
\nu_1 & = & +\nu_e \cos \delta + \nu_\mu \sin \delta \\ 
&&\\
\nu_2 & = & -\nu_e \sin \delta + \nu_\mu \cos \delta \;"
\end{array}
\end{equation}
\end{quote}

The authors introduced an interaction non-diagonal in $\nu_e$ and
$\nu_\mu$ (if mixing is assumed): 

\begin{equation}
{\cal L} = g \overline{\nu}_2
\nu_2 X^+ X
\end{equation}

\noindent where $g$ is a constant and $X$ is the field of heavy
particles.

Different effects of neutrino masses and mixing including
$\nu_\mu \rightarrow \nu_e$ transitions were
considered in the paper~\cite{MA}.

The authors shortly discussed the possibility to see effects of
 $\nu_\mu\rightarrow\nu_e$ 
transitions 
in Brookhaven neutrino experiment~\cite{DGGL},
aiming to check whether $\nu_e$ and $\nu_\mu$ were identical or
different particles, that at that time was going on.

I report below the part of
the paper~\cite{MA} dedicated to the consideration of  
$\nu_\mu\rightarrow\nu_e$ transitions:

\begin{quote}
``In the present case, however, weak neutrinos are {\em not stable} due to
occurrence of virtual transmutation $\nu_\mu \rightarrow \nu_e$ induced by
the interaction (4). If the mass difference between $\nu_1$ and
$\nu_2$, i.e. $|m_{\nu_2} - m_{\nu_1}| = m_{\nu_2} $, is assumed to be a
few
MeV the transmutation time $T\left( \nu_e \leftrightarrow \nu_\mu
\right)^2$ becomes $\sim$ $10^{-18}$ sec for fast neutrinos with momentum
$\sim$ $BeV/c$.

Therefore a chain of reactions such as

\begin{eqnarray}
\pi^+ & \rightarrow & \mu^+ \nu_\mu  \nonumber \\
\nu_\mu + Z\left(nucleus \right) & \rightarrow & Z^\prime +
      \left( \mu^- \; / \; e^- \right)
\end{eqnarray}

\noindent is useful to check the two-neutrino hypothesis only when
$|m_{\nu_2} -
m_{\nu_1}| \leq 10^{-6}$ MeV under the conventional geometry of the
experiments .
Conversely, the absence
of $e^-$ in the reaction (5) will be able
not only to verify two-neutrino hypothesis but also to provide an upper
limit of the mass of the second neutrino $\nu_2$ if the present scheme
should be accepted.''
\end{quote}

\section{ B. Pontecorvo 1967} 
B. Pontecorvo discussed all the possibilities of neutrino oscillations
in the case of two flavour neutrinos $\nu_e$ and $\nu_\mu$ in 
the paper~\cite{PO3}. In this paper he introduced the notion of sterile
neutrino:

\begin{quote}
``If there are two different additive lepton charges, the transitions
$\nu_e \leftrightarrow \overline{\nu}_e$ and $\nu_\mu \leftrightarrow
\overline{\nu}_\mu$ transform potentially {\em active} particles into
particles, which from the point of view of ordinary weak processes, are
sterile, {\em i.e.} practically undetectable, inasmuch as they have {\em
wrong} spirality. In such a case the only way of observing the effects
in question consists in measuring the intensity and the time variation
of the intensity of original active particles, but not in detecting the
appearance of the corresponding (sterile) antiparticles.''\footnote{At
that time Neutral Current processes were not known}
\end{quote}

B. Pontecorvo considered as a very favourable the
Zeldovich-Konopinsky-Mahmoud
scheme with  one lepton number for $e$ and $\mu$ and one four-component
neutrino.
In this scheme $\nu_L = \nu_e$, $\overline{\nu}_L = \nu_\mu$,
$\overline{\nu}_R = \overline{\nu}_e$ and $\nu_R = \overline{\nu}_\mu$
and there are no sterile neutrinos:

\begin{quote}
``Let us consider now the case when there is only one additive leptonic
charge the sign of which differs for $\mu^-$ and $e^-$. The proper
notation for the four neutral objects in such a case is $\nu_{left}$,
$\overline{\nu}_{left}$, $\nu_{right}$, $\overline{\nu}_{right}$. Then
the transitions $\nu_{left} \rightarrow \overline{\nu}_{left}$,
$\overline{\nu}_{right} \rightarrow \nu_{right}$ generate nonsterile
particles.

Returning to the usual notations, there will take place oscillations
$elneutrino \leftrightarrow mu$-$neutrino$\footnote{Old notations for
$\nu_e$ and $\nu_\mu$.}, which, in principle are detectable not only
by measuring the intensity and the time variation of the intensity of
original particles, but also by observing the appearance of new
particles.''
\end{quote}

In the same paper~\cite{PO3} oscillations of solar neutrinos were
considered for the first time.{\em Before the first results of Davis
experiment appeared}, B. Pontecorvo pointed out that due to neutrino
oscillations the observed flux of solar neutrinos could be twice smaller
than the expected flux. He anticipated the solar neutrino problem:

\begin{quote}
``From an observational point of view the ideal object is the sun. If
the oscillation length is smaller than the radius of the sun region
effectively producing neutrinos, (let us say one tenth of the sun radius
$R_\odot$ or 0.1 million km for $^8B$ neutrinos, which will give the
main contribution in the experiments being planned now), direct
oscillations will be smeared out and unobservable. The only effect on
the earth's surface would be that the flux of observable sun neutrinos
must be two times smaller than the total (active and sterile) neutrino
flux.''
\end{quote}

In this paper B. Pontecorvo discussed also the possible seasonal
variation of the flux of solar neutrinos that could appear in the case
if oscillation length is larger than the region
of the sun where neutrinos are produced. This possibility was pointed out
to him by I. Pomeranchuk.

\section{ V. Gribov, B. Pontecorvo 1969}

In the paper~\cite{GRPO} it was pointed out that
\begin{quote}
``solar neutrino oscillations is the most sensitive way of investigating
the question of lepton charge conservation.''
\end{quote}

In this paper the Majorana mass term, in which only left-handed fields
$\nu_{eL}$ and $\nu_{\mu L}$ enter, was for the first time considered,
and a scheme of neutrino mixing in which there are no transitions of
$\nu_e$ and $\nu_\mu$ into sterile states was proposed. Neutrinos with
definite masses $m_1$ and $m_2$ are in this scheme Majorana particles:

\begin{quote}
``The two component spinors $\nu_e$ and $\nu_\mu$ are not describing
anymore particles with zero mass but must be expressed in terms of
four-component Majorana spinors $\phi_1$ and $\phi_2$

\begin{equation}
\begin{array}{lll}
\nu_e & = & \frac{1}{2} \left(1 + \gamma_5 \right)
\left(+\phi_1 \cos \xi + \phi_2 \sin \xi \right)\\ 
&&\\
\nu_\mu & = & \frac{1}{2} \left( 1 + \gamma_5 \right) \left(
-\phi_1 \sin \xi + \phi_2 \cos \xi \right) \; "
\end{array}
\end{equation}
\end{quote}

Here

\begin{equation}
\tan 2\xi = \frac{2m_{e \overline{\mu}}}
{m_{\mu \overline{\mu}} - m_{e \overline{e}}}
\end{equation}

\noindent where $m_{e \overline{\mu}}$, $m_{\mu \overline{\mu}}$ and
$m_{e \overline{e}}$ are parameters that characterise the neutrino mass
matrix.

In the paper~\cite{GRPO} neutrino oscillations were considered:
\begin{quote}
``The mass difference between Majorana neutrinos described by $\phi_1$
and $\phi_2$ leads to oscillations $\nu_e \leftrightarrow \nu_\mu$,
$\overline{\nu}_e \leftrightarrow \overline{\nu}_\mu$. If at the time $t
= 0$, one electron neutrino is generated, the probability of observing
it at the time $t$ is

\begin{equation}
|\nu_e\left(t \right)|^2 = |\nu_e \left( 0 \right)|^2
\left\{
\frac{ m^2_- + 2 m^2_{e \overline{\mu}}}
{m^2_- + 4 m^2_{e \overline{\mu}}} +
\frac{2 m^2_{e \overline{\mu}}}
{m^2_- + 4 m^2_{e \overline{\mu}}} \cos \Delta t
\right\} 
\end{equation}

\noindent where

\begin{equation}
m_- = m_{e \overline{e}} - m_{\mu \overline{\mu}} 
\end{equation}

\begin{equation}
\Delta = \frac{1}{2p} \left(m^2_1 - m^2_2 \right) =
\frac{ m_{e \overline{e}} + m_{\mu \overline{\mu}} }{2p}
\sqrt{ m^2_- + 4 m^2_{e \overline{\mu}}} 
\end{equation}

and $p$ is the neutrino momentum.''
\end{quote} 

If $m_{e \overline{e}}$, $m_{\mu \overline{\mu}}$, $\ll m_{e
\overline{\mu}}$, or $m_{e \overline{e}} = m_{\mu \overline{\mu}}$ the
mixing is maximal $\left( \xi = \pi / 4 \right)$ and neutrino
oscillations are analogous to $K^0 \leftrightarrow \overline{K}{^0}$
oscillations. The authors consider this possibility as the most simple
and attractive one.

In the paper~\cite{JBA} written soon after the V.Gribov and B.Pontecorvo
paper appeared, J.Bahcall and S.Frautschi considered vacuum oscillations
of solar
neutrinos in details. In this paper the importance of averaging over
the solar
neutrino energy spectrum was demonstrated and stressed.

\section{ Quark-lepton analogy and neutrino oscillations}

After the Cabibbo-GIM quark mixing was established in the beginning
of the seventies the main arguments for neutrino mixing were based on
quark-lepton analogy~\cite{BIL},~\cite{ER}. In the
Gribov-Pontecorvo scheme
massive
neutrinos
are very different from other fundamental fermions: whereas charged
leptons and quarks are four-component Dirac particles, massive neutrinos
are two-component Majorana particles. In the paper~\cite{BIL}

\begin{quote}
``\ldots we consider neutrino mixing starting from a different point of
view suggested by an analogy between leptons and quarks. We assume that
each neutrino is described by a four-component spinor.''
\end{quote}

We stressed in~\cite{BIL} that the value of the lepton mixing
angle and 
the value of the Cabibbo angle could be completely different. We tried
to present different quantitative arguments that maximal mixing is the
most plausible and fruitful assumption. In the paper~\cite{BIPO}
we wrote:

\begin{quote}
``\ldots it seems to us that the special values of mixing angle $\theta
= 0$ (the usual scheme in which muonic charge is strictly conserved) and
$\theta = \pi /4$ are of the greatest interest\footnote{Today it looks like
small and $\pi/4$ values of neutrino mixing angles are really the
preferable
ones. Indeed in the framework of the mixing of three massive neutrinos
with three mixing angles $\theta_{12}$, $\theta_{13}$, $\theta_{23}$ from
CHOOZ and atmospheric neutrino experiments it follows that $\theta_{13}$
is small. From Super-Kamiokande atmospheric neutrino experiments it
follows that $\theta_{23}\simeq \pi/4$. The best fit of the
latest Super-Kamiokande
solar neutrino data was obtained for vacuum solution and large mixing
angle MSW solution with $\theta_{12} \simeq \pi/4$}
"
\end{quote}

For arbitrary angle $\theta$ the formulas for the oscillation of two types
of neutrinos, that are now standard, were presented in~\cite{BIPO2} 
and~\cite{PO4} :

\begin{equation}
\begin{array}{lllll}
I_{\nu_\ell}\left( R,p \right) & = &  \left[ 1- \frac{1}{2} \sin^2 2\theta
\left( 1- \cos 2\pi\frac{R}{L} \right) \right]
I^0_{\nu_\ell} \left( R,p \right) &  {\ell = e, \mu} & \\
& & & & \\
I_{\nu_{\ell^ \prime}} \left( R,p \right) & = & \left[
\frac{1}{2} \sin^2 2\theta \left( 1- \cos 2\pi\frac{R}{L} \right) \right]
I^0_{\nu_\ell} \left( R,p \right) & {\ell \neq \ell ^\prime,} &
\ell, \ell ^\prime = e, \mu
\end{array} 
\end{equation}

\noindent
Here $I_{\nu_ \ell}\left(
R,p \right)$, $I_{\nu_{\ell ^\prime}}\left( R,p \right)$ are the
intensities of $\nu_\ell$, $\nu_{\ell ^\prime}$ respectively, with
momentum $p$ at a distance $R$ from a source of $\nu_\ell$ neutrinos.
$I^0_{\nu_\ell} \left( R,p \right)$ is the intensity of $\nu_\ell$
neutrinos which would be expected in the absence of oscillations and 

\begin{equation}
L= \frac{4\pi p}{|m^2_2 - m^2_1|}
\end{equation}

\noindent
is the oscillation length. 

The list of papers on neutrino oscillations
published before 1977 is very
short:~\cite{PO2:57,MA,PO3}$~\bar{}$~\cite{BP:76}. In
our review~\cite{BIPO3} where we summarised the
status of neutrino oscillations and of other processes in which neutrino
masses and mixing can be revealed there are about seventy references. At
that time the idea of massless strictly two-component neutrinos still
prevailed.

The situation drastically changed after the appearance of Grand
Unification models and of the famous see-saw mechanism~\cite{GELY} that
connect
the smallness of neutrino masses with the violation of the lepton
numbers at very large scale. From the point of view of the Grand
Unification models it is natural and plausible that neutrinos have a mass
and the
investigation of the neutrino masses and mixing can provide a probe for a
new scale in physics. At the end of the seventies new experiments
specially dedicated to search for neutrino oscillations were started.
In our review on neutrino masses and oscillations~\cite{BIPET}, written
in 1986, there
are about 230 references.

The real massive neutrino boom started after the Super-Kamiokande
result was announced at the {\em Neutrino 98} conference at Takayama.
More than 450 papers on the neutrino mass, mixing and oscillations
appeared in the HEP archive in less than one year.

\section{ Conclusion}

There is at present a convincing evidence that neutrinos are
massive and mixed particles. This evidence has been obtained from
atmospheric neutrino experiments (Kamiokande~\cite{FU}, IMB~\cite{BES}, 
Soudan 2~\cite{AL} and MACRO~\cite{AM}), and first of all from the
Super-Kamiokande
experiment~\cite{KF} and from all solar neutrino experiments 
(Homestake~\cite{DCL}, Gallex~\cite{WH}, Sage~\cite{AB}, Kamiokande~\cite{HI} 
and Super-Kamiokande~\cite{YSU}). From the theoretical point of view it is
very plausible that  neutrinos have a mass possibly connected with a new
scale in physics.

It required many years of work and heroic efforts of many
experimental groups to reveal effects of tiny neutrino masses. It is
difficult not
to give tribute to the great intuition of B. Pontecorvo who pursued the
idea of neutrino oscillations for many years at a time when the
general opinion, mainly based on the success of the two-component
neutrino theory, favoured massless neutrinos \footnote{For a collection 
of the papers of B. Pontecorvo see~\cite{BOOK}.}.

From my own point of view the history of the neutrino oscillations is an
illustration of the importance of analogy in physics. It is also an
illustration of the importance of new courageous ideas not always in
agreement with the general opinion.

\section*{Acknowledgement}
It is a pleasure for me to express my gratitude
to G.Fidecaro
for very useful discussion

\end{document}